\documentclass[11pt]{article}
\usepackage[centertags]{amsmath}
\usepackage{amsfonts}
\usepackage{amssymb}
\usepackage{amsthm}
\usepackage{newlfont}
\usepackage{graphicx}

\newfont{\bb}{msbm10 at 12pt}

\begin{document}

\title{New Class of Magnetized Inhomogeneous Bianchi Type-I Cosmological Model with Variable Magnetic Permeability in Lyra Geometry}
\author{ Ahmad T Ali$^{1,2}$ and F Rahaman$^{3}$\\
$1$- King Abdul Aziz University,\\
Faculty of Science, Department of Mathematics,\\
PO Box 80203, Jeddah, 21589, Saudi Arabia.\\
$2$- Mathematics Department,\\
Faculty of Science, Al-Azhar University,\\
Nasr City, 11884, Cairo, Egypt.\\
$3$-Department of Mathematics,\\ Jadavpur University, Kolkata 700
032, West Bengal,
India\\
E-mail: atali71@yahoo.com and  rahaman@iucaa.ernet.in}

\date{\today}

\maketitle
\begin{abstract} Inhomogeneous Bianchi type-I cosmological model
with electro-magnetic field based on Lyra geometry is investigated. Using separated method, the Einstein
 field equations have been solved
analytically with the aid of {\it{Mathematica}} programm. A new class of exact solutions
 have been obtained by considering the potentials of metric and displacement field are functions of
  coordinates $t$ and $x$. We have assumed that $F_{12}$ is the only non-vanishing component of electro-magnetic
field tensor $F_{ij}$ . The Maxwell's equations show that $F_{12}$ is the function
of $x$ alone whereas the magnetic permeability is the function of $x$ and $t$
both. To get the deterministic solution, it has been assumed that the
expansion scaler $\Theta$ in the model is proportional to the value $\sigma_1^1$ of the
shear tensor $\sigma_i^j$ . Some physical and geometric properties of the model
are also discussed and graphed.
\end{abstract}



\section{Introduction }

The study of  cosmology is to know the large scale structure of the Universe. The observations indicate that the present universe is not exactly spatially
homogeneous.  Also, the inhomogeneity plays a crucial role in the process of structure formation, especially in the
context of galaxy formation. Therefore, it will be interesting to study inhomogeneous cosmological models. The inhomogeneous
cosmological models help to provide information in the understanding of the formation of galaxies during the early stages
of their evolution.

The best known inhomogeneous cosmological model is the Lemaître-Tolman model (or LT model) which deals with  the study of structure in
the universe by means of exact solutions of Einstein's field equations. Some other known exact solutions of inhomogeneous
cosmological models are the Szekeres metric, Szafron metric, Stephani metric, Kantowski-Sachs metric, Barnes metric, Kustaanheimo-Qvist metric and Senovilla metric \cite{AK}.

Zel'dovich \cite{zel} argued that various astrophysical phenomena lead to the existence of  magnetic fields in the galactic and intergalactic
spaces. Also Harrison \cite{ha} has suggested that there is a close connection of magnetic field with the  cosmological origin. Melvin \cite{mel}
suggested that at the early stages of its evolution when the
universe underwent several phase transition,  the matter was in a
highly ionized state and was smoothly coupled with the field.
During the expansion of the early universe, after the Planck time,
ions were combined to form neutral matter.
 Hence the presence of
magnetic field in   the energy- momentum tensor of the early
universe is not unrealistic.

Cylindrically-symmetric space-time is more general than the
Robertson-Walker spherically symmetric space-time and plays an
important role in the study of  the universe when the anisotropy
and inhomogeneity are taking into consideration.

Einstein's general
    theory
    relativity is based on Riemannain geometry. If one modifies the Riemannian geometry, then Einstein's field
equations will be changed automatically from its original form.
 Modification of Riemannian geometry   have developed to
solve the problems such as unification of gravitation with
electromagnetism, problems arising when the gravitational field is
coupled to matter fields, singularities of standard cosmology etc.
In recent years there has been considerable interest in
alternative theory of
   gravitation to explain the above unsolved problems.
Long ago, since 1951, Lyra  \cite{ly} proposed a modification of Riemannian
geometry
    by introducing a gauge function into the structure-less manifold that bears a close
    resemblances to Weyl's geometry.

Using the above modification of Riemannian geometry Sen \cite{sen1} and
Sen and Dunn \cite{SD}  proposed a new scalar tensor
    theory of gravitation and constructed very similar
to Einstein field equations.  Based on
      Lyra's geometry the field equations can be written as \cite{sen1}

\begin{equation}\label{Eq1}
R_{ij}-\frac{1}{2}\,g_{ij}\,R+\frac{3}{2}\,\phi_i\,\phi_j-\frac{3}{4}\,g_{ij}\,\phi_k\,\phi^k=-8\,\pi\,G\,T_{ij},
\end{equation}

where $\phi_a$ is the displacement vector and other symbols have their usual meaning as in Riemannian geometry.

Halford \cite{hal} has argued that the nature  of  constant displacement field  $\phi_i$ in Lyra's geometry  is very similar to cosmological constant $\Lambda$ in the normal general relativistic theory. Halford also predicted that the present theory will provide the same effects within observational limits, as far as the classical solar system tests are concerned, as well as tests based on the linearized form of field equations. For a review on Lyra Geometry, one can see the reference \cite{book}.

Recently, Pradhan et al. \cite{prad21, prad22, prad23, prad23-2, prad23-3, prad23-4, prad24}, Casama et al. \cite{casama1}, Rahaman et al. \cite{rahaman1}, Bali and
Chandnani \cite{bali11, bali12}, Kumar and Singh \cite{kumar1}, Yadav et al. \cite{yadav1}, Rao et al. \cite{rao1}, Zia and Singh \cite{zia1} have studied cosmological models based on Lyra's geometry in various contexts.

In this work, we attempted to find a new class of exact cosmological solutions for the universe.
Here, we investigate inhomogeneous Bianchi type-I cosmological model with electro-magnetic field based on
 Lyra geometry.

The outline of the paper is as follows: The metric and the field equations are presented in section 2.
 In section 3 we found new class of exact solutions for  the modified Einstein field equations.
      Section 4 discusses some physical and geometrical properties of the obtained model.
The last section 5 contains
     concluding remarks about the proposal.

\section{The metric and field equations}

We consider Bianchi type-I metric, with the convention $(x^0=t,\,x^1=x,\,x^2=y,\,x^3=z)$, in the form \cite{bali13, bali14, katore1, saman1}
\begin{equation}  \label{u21}
ds^2=dt^2-A^2\,dx^2-B^2\,dy^2-C^2\,dz^2,
\end{equation}
where $A$ is a function of $t$ only while $B$ and $C$ are functions of $x$ and $t$. The proper volume of the model (\ref{u21}) is given by
\begin{equation}  \label{u22-1}
V=\sqrt{-g}=A\,B\,C.
\end{equation}
The four-acceleration vector, the rotation, the expansion scalar and the shear scalar characterizing the
four velocity vector field, $u^i$, which satisfying the relation in co-moving coordinate system
\begin{equation}  \label{u22-2}
g_{ij}\,u^i\,u^j=1 \,\,\, \mathrm{and} \,\,\, u^i=u_i=(1,0,0,0).
\end{equation}
respectively, have the usual definitions as given by Raychaudhuri \cite{rayc1}
\begin{equation}  \label{u22-3}
\left\{
  \begin{array}{ll}
    \dot{u}_i\,&=\,u_{i;j}\,u^j,\\
   \omega_{ij}\,&=\,u_{[i;j]}+\dot{u}_{[i}\,u_{j]},\\
  \Theta\,&=\,u_{;i}^{i},\\
 \sigma^2\,&=\,\frac{1}{2}\,\sigma_{ij}\,\sigma^{ij},
   \end{array}
\right.
\end{equation}
where
\begin{equation}  \label{u22-4}
  \begin{array}{ll}
\sigma_{ij}=u_{(i;j)}+\dot{u}_{(i}\,u_{j)}-\frac{1}{3}\Theta(g_{ij}+u_i\,u_j).
\end{array}
\end{equation}
In view of the metric (\ref{u21}), the four-acceleration vector, the rotation, the expansion scaler and
the shear scalar given by (\ref{u22-3}) can be written in a co-moving coordinates system as
\begin{equation}  \label{u22-5}
\left\{
  \begin{array}{ll}
    \dot{u}_i\,&=\,0,\\
   \omega_{ij}\,&=\,0,\\
  \Theta\,&=\,\dfrac{\dot{A}}{A}+\dfrac{B_t}{B}+\dfrac{C_t}{C},\\
 \sigma^2\,&=\,\dfrac{5\,\dot{A}^2}{9\,A^2}+\dfrac{\dot{A}\,B_t}{9\,A\,B}+\dfrac{5B^2_t}{9\,B^2}+\dfrac{\dot{A}\,C_t}{9\,A\,C}++\dfrac{B_t\,C_t}{9\,B\,C}
+\dfrac{5C^2_t}{9\,C^2},
   \end{array}
\right.
\end{equation}
where the non-vanishing components of the shear tensor $\sigma_i^j$ are
\begin{equation}  \label{u22-6}
\left\{
  \begin{array}{ll}
    \sigma_1^1\,&=\,\dfrac{2\,\dot{A}}{3\,A}-\dfrac{B_t}{3\,B}-\dfrac{C_t}{3\,C},\\
   \sigma_2^2\,&=\,\dfrac{2B_t}{3\,B}-\dfrac{\dot{A}}{3\,A}-\dfrac{C_t}{3\,C},\\
  \sigma_3^3\,&=\,\dfrac{2\,C_t}{3\,C}-\dfrac{\dot{A}}{3\,A}-\dfrac{B_t}{3\,B},\\
 \sigma_4^4\,&=\,-\dfrac{2}{3}\Big(\dfrac{\dot{A}}{A}+\dfrac{B_t}{B}+\dfrac{C_t}{C}\Big).
   \end{array}
\right.
\end{equation}

To study the cosmological model, we use the  field equations in Lyra geometry given in (1) in which
the displacement field vector $\phi_i$ is given  by
\begin{equation}  \label{u24}
\phi_i=\big(\beta(x,t),0,0,0\big).
\end{equation}
$T_{ij}$ is the energy momentum tensor given by
\begin{equation}  \label{u25}
T_{ij}=(\rho+p)u_iu_j-pg_{ij}+E_{ij},
\end{equation}
where $E_{ij}$ is the electro-magnetic field given by Lichnerowicz \cite{lich1}:
\begin{equation}  \label{u26}
E_{ij}=\bar{\mu}\big[h_lh^l\big(u_iu_j-\frac{1}{2}g_{ij}\big)+h_ih_j\big].
\end{equation}
Here $\rho$ and $p$ are the energy density and isotropic pressure, respectively while $\bar{\mu}$ is
the magnetic permeability and $h_i$ the magnetic flux vector defined by:
\begin{equation}  \label{u27}
h_i=\dfrac{\sqrt{-g}}{2\bar{\mu}}\epsilon_{ijkl}F^{kl}u^j.
\end{equation}
$F_{ij}$ is the electro-magnetic field tensor and $\epsilon_{ijkl}$ is a Levi-Civita tensor density.
If we consider the current flow along $z$-axis, then $F_{12}$ is only non-vanishing component of $F_{ij}$. Then the Maxwell's equations
\begin{equation}  \label{u28}
F_{ij;k}+F_{jk;i}+F_{ki;j}=0
\end{equation}
and
\begin{equation}  \label{u29}
\Big[\dfrac{1}{\bar{\mu}}F^{ij}\Big]_{;j}=J^i
\end{equation}
require that $F_{12}$ be function of $x$ alone \cite{prad1}. We assume that the magnetic permeability as
 a function of both $x$ and $t$. Here the semicolon represents a covariant differentiation.\\

For the line element  (\ref{u21}) the field equation (1) can be reduced to the following system
of non-linear partial differential equations:
\begin{equation}  \label{u210}
\begin{array}{ll}
    E_1=\dfrac{B_{xt}}{B}+\dfrac{C_{xt}}{C}-\dfrac{\dot{A}}{A^2}\Big(\dfrac{B_x}{B}+\dfrac{C_x}{C}\Big)=0,
  \end{array}
\end{equation}

\begin{equation}  \label{u211}
\begin{array}{ll}
        E_2=\dfrac{B_{tt}}{B}+\dfrac{B_t\,C_t}{B\,C}+\dfrac{1}{A^2}\Big(\dfrac{C_{xx}}{C}-\dfrac{B_x\,C_x}{B\,C}\Big)
-\dfrac{\dot{A}\,C_t}{A\,C}-\dfrac{\ddot{A}}{A}=0,
  \end{array}
\end{equation}

\begin{equation}  \label{u212}
\begin{array}{ll}
        \chi\,\rho+\dfrac{3}{4}\beta^2=\dfrac{C_{tt}}{2\,C}+\dfrac{3\,B_t\,C_{t}}{2\,B\,C}
           -\dfrac{1}{A^2}\Big(\dfrac{C_{xx}}{C}+\dfrac{B_{xx}}{2\,B}+\dfrac{3\,B_x\,C_x}{2\,B\,C}\Big)
        +\dfrac{\dot{A}}{A}\Big(\dfrac{B_t}{2\,B}+\dfrac{C_t}{C}\Big)-\dfrac{\ddot{A}}{A},
  \end{array}
\end{equation}

\begin{equation}  \label{u213}
\begin{array}{ll}
        \chi\,p+\dfrac{3}{4}\beta^2=\dfrac{1}{A^2}\Big(\dfrac{B_{xx}}{2\,B}+\dfrac{B_x\,C_x}{2\,A\,B}\Big)
        -\dfrac{B_{tt}}{B}-\dfrac{C_{tt}}{2\,C}-\dfrac{B_t\,C_t}{2\,B\,C}
        -\dfrac{\dot{A\,B_t}}{2\,A\,B}-\dfrac{\ddot{A}}{2\,A},
  \end{array}
\end{equation}

\begin{equation}  \label{u214}
\begin{array}{ll}
        \dfrac{\chi\,F^2_{12}}{\bar{\mu}\,A^2\,B^2}=\dfrac{C_{tt}}{C}+\dfrac{B_t\,C_t}{B\,C}+\dfrac{1}{A^2}\Big(\dfrac{B_{xx}}{B}-\dfrac{B_x\,C_x}{B\,C}\Big)
-\dfrac{\dot{A}\,B_t}{A\,B}-\dfrac{\ddot{A}}{A},
  \end{array}
\end{equation}
where $\chi=8\,\pi\,G$.

\section{Solutions of the field equations}

The field equations (\ref{u210})-(\ref{u214}) constitute a system of five highly non-linear differential equations
 with seven unknowns variables, $A$, $B$, $C$, $p$, $\rho$, $F^2_{12}/\mu$ and $\beta$. Therefore, two physically reasonable
  conditions amongst these parameters are required to obtain explicit solutions of the field equations. First, Let us
   assume that the density $\rho$ and the pressure $p$ are related by baro-tropic equation of state:
\begin{equation}  \label{u31}
\begin{array}{ll}
p=\lambda\,\rho, \,\,\,\,\, 0\,\leq\,\lambda\leq\,1.
\end{array}
\end{equation}
The second required condition is by assuming that the expansion scalar $\Theta$ in the model (\ref{u21}) is
 proportional to the eigenvalue $\sigma_1^1$ of the shear tensor $\sigma_j^k$. Then from (\ref{u22-5}) and (\ref{u22-6}), we get
\begin{equation}\label{u33}
  \begin{array}{ll}
\dfrac{1}{3}\Big(\dfrac{2\,\dot{A}}{A}-\dfrac{B_t}{B}-\dfrac{C_t}{C}\Big)=\dfrac{c_1}{3}\Big(\dfrac{\dot{A}}{A}+\dfrac{B_t}{B}+\dfrac{C_t}{C}\Big),
  \end{array}
\end{equation}
where $\dfrac{c_1}{3}$ is a constant of proportionality. Hence, the above condition can be written in the form
\begin{equation}\label{u34}
  \begin{array}{ll}
\dfrac{\big(B\,C\big)_t}{B\,C}=\Big(\dfrac{2-c_1}{1+c_1}\Big)\dfrac{\dot{A}}{A},
  \end{array}
\end{equation}
By integration the above equation with respect to $t$, we get:
\begin{equation}\label{u35}
  \begin{array}{ll}
B\,C=f(x)\,A^{a_1},
  \end{array}
\end{equation}
where $a_1=\dfrac{2-c_1}{1+c_1}$ and the constant of integration here is $f$ function of $x$. Now, we can take the following assumption:
\begin{equation}\label{u36}
B(x,t)=f(x)\,k(x)\,l(t).
\end{equation}
If we substitute the assumptions in equation (\ref{u36}), into (\ref{u210}), we have the following condition:
\begin{equation}\label{u37}
\dfrac{A(t)\,\dot{l}(t)-l(t)\,\dot{A}(t)}{a_1\,l(t)\,\dot{A}(t)-2\,A(t)\,\dot{l}(t)}=\dfrac{f(x)\,k'(x)}{k(x)\,f'(x)}=c_2,
\end{equation}
where $c_2$ is an arbitrary constant. The above condition leads to
\begin{equation}\label{u38}
k(x)=c_3\,f^{c_2}(x),\,\,\,\,\,\,\,\,\,\,l(t)=c_4\,A^{a_2}(t),
\end{equation}
where $a_2=\dfrac{1+a_1c_2}{1+2c_2}$ while $c_3$ and $c_4$ are constants of integration. Therefore, the equation (\ref{u211}) leads to:
\begin{equation}\label{u39}
\dfrac{(2-a_1)\big[a_1\,\dot{A}^2(t)+A\,\ddot{A}\big]}{1+2c_2}=\dfrac{2(1+c_2)f'^{2}(x)-f(x)\,f''(x)}{f^2(x)}=\dfrac{c_5}{1+2c_2},
\end{equation}
where $c_5$ is an arbitrary constant. Solving the above ordinary differential equation of $f(x)$ we have
\begin{equation}\label{u310}
f(x)\,=\,\left\{
  \begin{array}{ll}
    \,\mathrm{cosh}^{a_3}[a_0\,x+c_6], & \,\,\,\,\,c_5=a_0^2>0, \\
\\
    \,\mathrm{cos}^{a_3}[a_0\,x+c_6],, & \,\,\,\,\,c_5=-a_0^2<0
  \end{array}
\right.
\end{equation}
where $a_3=-\dfrac{1}{1+2c_2}$ and $c_7$ is an arbitrary constant of integration, while another ordinary differential equation of $A(t)$, can be written as:
\begin{equation}\label{u311}
A(t)\,\ddot{A}(t)+a_1\,\dot{A}^2(t)=a_4,
\end{equation}
where $a_4=\dfrac{c_5}{2-a_1}$. If we integrate the above equation, we can get:
\begin{equation}\label{u312}
\dot{A}^2(t)=c_8\,A^{-2a_1}+a_5,
\end{equation}
where $a_5=\dfrac{a_4}{a_1}$ while $c_8$ is a constant of integration.

Then the general solution can be written in the following one of the form:
\begin{equation}\label{u313-1}
\mathrm{When}\,\,\, c_5>0,\,\,\,\,\,\,\left\{
  \begin{array}{ll}
A(t)\,\,\,\,\,\,\,\,\,\,\,\,\,\,\,\text{is satisfied the equation (\ref{u312})},\\
B(x,t)=a_6\,A^{a_2}(t)\,\mathrm{cosh}^{a_3-a_7}[a_0\,x+c_6],\\
C(x,t)=a_8\,A^{a_1-a_2}(t)\,\mathrm{cosh}^{a_7}[a_0\,x+c_6],
  \end{array}
\right.
\end{equation}
or
\begin{equation}\label{u313-2}
\mathrm{When}\,\,\,c_5<0,\,\,\,\,\,\,\left\{
  \begin{array}{ll}
A(t)\,\,\,\,\,\,\,\,\,\,\,\,\,\,\,\text{is satisfied the equation (\ref{u312})},\\
B(x,t)=a_6\,A^{a_2}(t)\,\mathrm{cos}^{a_3-a_7}[a_0\,x+c_6],\\
C(x,t)=a_8\,A^{a_1-a_2}(t)\,\mathrm{cos}^{a_7}[a_0\,x+c_6],
  \end{array}
\right.
\end{equation}
where $a_6=c_3\,c_4\,c_7^{1+c_2}$, $a_7=\dfrac{c_2}{1+2c_2}$ and $a_8=\dfrac{1}{c_3\,c_4\,c_7^{c_2}}$.\\

Thus the line element with these coefficients can be written in the following general form:
\begin{equation}  \label{u314-1}
\begin{array}{ll}
ds^2=dt^2-A^2(t)\,dx^2-a_6^2\,A^{2\,a_2}(t)\,\mathrm{cosh}^{2a_3-2a_7}[a_0\,x+c_6]\,dy^2\\
\,\,\,\,\,\,\,\,\,\,\,\,\,\,\,\,\,\,\,\,\,\,\,\,\,\,\,\,\,\,\,\,\,\,\,\,\,\,\,\,\,\,\,\,\,\,\,\,\,\,\,\,\,\,\,\,\,\,\,\,
\,\,\,\,\,\,\,\,\,\,\,\,\,\,\,\,\,-a_8^2\,A^{2a_1-2a_2}(t)\,\mathrm{cosh}^{2\,a_7}[a_0\,x+c_6]\,dz^2,
\end{array}
\end{equation}
or
\begin{equation}  \label{u314-2}
\begin{array}{ll}
ds^2=dt^2-A^2(t)\,dx^2-a_6^2\,A^{2\,a_2}(t)\,\mathrm{cos}^{2a_3-2a_7}[a_0\,x+c_6]\,dy^2\\
\,\,\,\,\,\,\,\,\,\,\,\,\,\,\,\,\,\,\,\,\,\,\,\,\,\,\,\,\,\,\,\,\,\,\,\,\,\,\,\,\,\,\,\,\,\,\,\,\,\,\,\,\,\,\,\,\,\,\,\,
\,\,\,\,\,\,\,\,\,\,\,\,\,\,\,\,\,-a_8^2\,A^{2a_1-2a_2}(t)\,\mathrm{cos}^{2\,a_7}[a_0\,x+c_6]\,dz^2,
\end{array}
\end{equation}
where $a_2=\dfrac{a_1-2\,a_3+a_1\,a_3}{2}$, $a_5=\dfrac{c_5}{2a_1-a_1^2}$, $a_7=\dfrac{1+a_3}{2}$, $A(t)$ satisfied the
 equation (\ref{u312}) while $a_1$, $a_3$, $a_6$, $a_8$, $c_5$, $c_6$ and $c_8$ are arbitrary constants.\\

Now, for some special cases of the constants $a_1$, $c_5$ and $c_8$, we can find a class of solutions of the model under
 study, as the following:\\

\textbf{Solution (1):} When $c_8=0$, the solution of equation (\ref{u312}) is:
\begin{equation}\label{u315-1}
\mathrm{When}\,\,\,a_0=d_1\,\sqrt{a_1\,(2-a_1)},\,\,\,\,\,\left\{
  \begin{array}{ll}
A(t)=d_1\big(t+b_1\big),\\
B(x,t)=m_1\,\big(t+b_1\big)^{a_2}\,\mathrm{cosh}^{a_3-a_7}[a_0\,x+c_6],\\
C(x,t)=n_1\,\big(t+b_1\big)^{a_1-a_2}\,\mathrm{cosh}^{a_7}[a_0\,x+c_6],
  \end{array}
\right.
\end{equation}
or
\begin{equation}\label{u315-2}
\mathrm{When}\,\,\,a_0=d_1\,\sqrt{a_1\,(a_1-2)},\,\,\,\,\,\left\{
  \begin{array}{ll}
A(t)=d_1\big(t+b_1\big),\\
B(x,t)=m_1\,\big(t+b_1\big)^{a_2}\,\mathrm{cos}^{a_3-a_7}[a_0\,x+c_6],\\
C(x,t)=n_1\,\big(t+b_1\big)^{a_1-a_2}\,\mathrm{cos}^{a_7}[a_0\,x+c_6],
  \end{array}
\right.
\end{equation}
where $m_1=a_6\,d_1^{a_2}$, $n_1=a_8\,d_1^{a_1-a_2}$, $d_1$, $b_1$,  $a_1$, $a_3$ and $c_6$ are arbitrary constants.\\

\textbf{Solution (2):} When $a_1=1$, then the solution of equation (\ref{u312}) is:
\begin{equation}\label{u316-1}
\mathrm{When}\,\,\,c_5=a_0^2,\,\,\,\,\,\left\{
  \begin{array}{ll}
A(t)=a_0\,\sqrt{\big(t+b_2\big)^2+r_2},\\
B(x,t)=m_2\,\Big[\big(t+b_2\big)^2+r_2\Big]^{\dfrac{1-a_3}{4}}\,\mathrm{cosh}^{\dfrac{a_3-1}{2}}[a_0\,x+c_6],\\
C(x,t)=n_2\,\Big[\big(t+b_2\big)^2+r_2\Big]^{\dfrac{1+a_3}{4}}\,\mathrm{cosh}^{\dfrac{a_3+1}{2}}[a_0\,x+c_6],,
  \end{array}
\right.
\end{equation}
or
\begin{equation}\label{u316-2}
\mathrm{When}\,\,\,c_5=-a_0^2,\,\,\,\,\,\left\{
  \begin{array}{ll}
A(t)=a_0\,\sqrt{\tilde{r}_2-\big(t+b_2\big)^2},\\
B(x,t)=m_2\,\Big[\tilde{r}_2-\big(t+b_2\big)^2\Big]^{\dfrac{1-a_3}{4}}\,\mathrm{cos}^{\dfrac{a_3-1}{2}}[a_0\,x+c_6],\\
C(x,t)=n_2\,\Big[\tilde{r}_2-\big(t+b_2\big)^2\Big]^{\dfrac{1+a_3}{4}}\,\mathrm{cos}^{\dfrac{a_3+1}{2}}[a_0\,x+c_6],,
  \end{array}
\right.
\end{equation}
where $r_2=-\dfrac{c_8}{a_0^4}$, $\tilde{r}_2=\dfrac{c_8}{a_0^4}\,>\,0$, $m_2=a_6\,a_0^{\dfrac{1-a_3}{2}}$, $n_2=a_8\,
a_0^{\dfrac{1+a_3}{2}}$,  $b_2$,  $a_3$, $a_0$ and $c_6$ are arbitrary constants.\\

\textbf{Solution (3):} When $a_1=-1$ then $c_8=\tilde{a}_0^2\,>\,0$ and the solution of equation (\ref{u312}) is:
\begin{equation}\label{u317-1}
\mathrm{When}\,\,c_5=a_0^2,\,\,\left\{
  \begin{array}{ll}
A(t)=\dfrac{1}{4\,\tilde{a}_0^2}\,\mathrm{exp}\big[\tilde{a}_0\,(t+b_3)\big]+\dfrac{a_0^2}{3}\,\mathrm{exp}\big[-\tilde{a}_0\,(t+b_3)\big],\\
B(x,t)=a_6\,\mathrm{cosh}^{\dfrac{a_3-1}{2}}[a_0\,x+c_6]\times\\
\Big[\dfrac{1}{4\,\tilde{a}_0^2}\,\mathrm{exp}\big[\tilde{a}_0\,(t+b_3)\big]+\dfrac{a_0^2}{3}\,\mathrm{exp}\big[-\tilde{a}_0\,(t+b_3)\big]
\Big]^{\dfrac{-3a_3-1}{2}},\\
C(x,t)=a_8\,\mathrm{cosh}^{\dfrac{a_3+1}{2}}[a_0\,x+c_6]\times\\
\Big[\dfrac{1}{4\,\tilde{a}_0^2}\,\mathrm{exp}\big[\tilde{a}_0\,(t+b_3)\big]+\dfrac{a_0^2}{3}\,\mathrm{exp}\big[-\tilde{a}_0\,(t+b_3)\big]
\Big]^{\dfrac{3a_3-1}{2}},
  \end{array}
\right.
\end{equation}
or
\begin{equation}\label{u317-2}
\mathrm{When}\,\,c_5=-a_0^2,\,\,\left\{
  \begin{array}{ll}
A(t)=\dfrac{1}{4\,\tilde{a}_0^2}\,\mathrm{exp}\big[\tilde{a}_0\,(t+b_3)\big]-\dfrac{a_0^2}{3}\,\mathrm{exp}\big[-\tilde{a}_0\,(t+b_3)\big],\\
B(x,t)=a_6\,\mathrm{cos}^{\dfrac{a_3-1}{2}}[a_0\,x+c_6]\times\\
\Big[\dfrac{1}{4\,\tilde{a}_0^2}\,\mathrm{exp}\big[\tilde{a}_0\,(t+b_3)\big]-\dfrac{a_0^2}{3}\,\mathrm{exp}\big[-\tilde{a}_0\,(t+b_3)\big]
\Big]^{\dfrac{-3a_3-1}{2}},\\
C(x,t)=a_8\,\mathrm{cos}^{\dfrac{a_3+1}{2}}[a_0\,x+c_6]\times\\
\Big[\dfrac{1}{4\,\tilde{a}_0^2}\,\mathrm{exp}\big[\tilde{a}_0\,(t+b_3)\big]-\dfrac{a_0^2}{3}\,\mathrm{exp}\big[-\tilde{a}_0\,(t+b_3)\big]
\Big]^{\dfrac{3a_3-1}{2}},
  \end{array}
\right.
\end{equation}
where $\tilde{a}_0$, $a_6$, $a_8$, $a_3$, $b_3$, $a_0$ and $c_6$ are arbitrary constants.\\

\textbf{Solution (4):} When $a_1=-\dfrac{1}{2}$, then the solution of equation (\ref{u312}) is:
\begin{equation}\label{u318}
\mathrm{When}\,c_5=a_0^2,\,\left\{
  \begin{array}{ll}
A(t)=d_4\Big[\big(t+b_4\big)^2+\dfrac{a_0^2}{5\,d_4^2}\Big],\\
B(x,t)=m_4\,\Big[\big(t+b_4\big)^2+\dfrac{a_0^2}{5\,d_4^2}\Big]^{\dfrac{-5\,a_3-1}{4}}\,\mathrm{cosh}^{\dfrac{a_3-1}{2}}[a_0\,x+c_6],\\
C(x,t)=n_4\,\Big[\big(t+b_4\big)^2+\dfrac{a_0^2}{5\,d_4^2}\Big]^{\dfrac{5\,a_3-1}{4}}\,\mathrm{cosh}^{\dfrac{a_3+1}{2}}[a_0\,x+c_6]
  \end{array}
\right.
\end{equation}
or
\begin{equation}\label{u318}
\mathrm{When}\,c_5=-a_0^2,\,\left\{
  \begin{array}{ll}
A(t)=d_4\Big[\big(t+b_4\big)^2-\dfrac{a_0^2}{5\,d_4^2}\Big],\\
B(x,t)=m_4\,\Big[\big(t+b_4\big)^2-\dfrac{a_0^2}{5\,d_4^2}\Big]^{\dfrac{-5\,a_3-1}{4}}\,\mathrm{cos}^{\dfrac{a_3-1}{2}}[a_0\,x+c_6],\\
C(x,t)=n_4\,\Big[\big(t+b_4\big)^2-\dfrac{a_0^2}{5\,d_4^2}\Big]^{\dfrac{5\,a_3-1}{4}}\,\mathrm{cos}^{\dfrac{a_3+1}{2}}[a_0\,x+c_6]
  \end{array}
\right.
\end{equation}
where $d_4=\dfrac{c_8}{4}$, $m_4=a_6\,d_4^{\dfrac{-5a_3-1}{4}}$, $n_4=a_8\,d_4^{\dfrac{5a_3-1}{4}}$, $a_3$, $b_4$, $a_0$ and $c_6$ are arbitrary constants.\\

\textbf{Solution (5):} When $a_1=-2$, then $c_5=-a_0^2$ and the solution of equation (\ref{u312}) is Jacobi elliptic function as follows:
\begin{equation}\label{u319}
\left\{
  \begin{array}{ll}
A(t)=q_1\,\mathrm{cn}\Big(\dfrac{a_0}{2q_1}\big(t+b_5\big);\dfrac{1}{2}\Big),\\
B(x,t)=m_5\,\mathrm{cos}^{\dfrac{a_3-1}{2}}[a_0\,x+c_6]\,\mathrm{cn}^{-2a_3-1}\Big(\dfrac{a_0}{2q_1}\big(t+b_5\big);\dfrac{1}{2}\Big),\\
C(x,t)=n_5\,\mathrm{cos}^{\dfrac{a_3+1}{2}}[a_0\,x+c_6]\,\mathrm{cn}^{2a_3-1}\Big(\dfrac{a_0}{2q_1}\big(t+b_5\big);\dfrac{1}{2}\Big),
  \end{array}
\right.
\end{equation}
where $m_5=a_6\,q_1^{-2a_3-1}$, $n_5=a_8\,q_1^{2a_3-1}$, $q_1$, $a_3$, $b_5$, $a_0$ and $c_6$ are arbitrary constants.\\

It is well known that \cite{ali1, elsa1, elsa2}, $\text{sn}{(\xi;m^2)}$,
$\text{cn}{(\xi;m^2)}$ and
$\text{dn}{(\xi;m^2)}$ are called the Jacobian
elliptic sine function, the Jacobian elliptic cosine function and
the Jacobian elliptic function of third kind respectively, and
$0<m^2<1$ is the modulus of the Jacobian elliptic function.

\section{Physical properties of the model}

Using equations (\ref{u313-1}) and (\ref{u313-2}) in the Einstein field equations (\ref{u212}) and (\ref{u214}), with take
 into account the condition (\ref{u31}), the expressions for density $\rho$, pressure $p$
  and displacement field $\beta$ are given by:\\

\textbf{For the model (\ref{u314-1})}

\begin{equation}\label{u52}
  \begin{array}{ll}
\rho(x,t)=\dfrac{a_1\,a_5}{\chi\,(1-\lambda)\,A^2(t)}\Bigg[a_1+a_3\,(a_1-2)\Big(a_3+(1-a_3)\,\mathrm{sech}^2[a_0\,x+c_6]\Big)\Bigg],
  \end{array}
\end{equation}

\begin{equation}\label{u53}
  \begin{array}{ll}
p(x,t)=\dfrac{\lambda\,a_1\,a_5}{\chi\,(1-\lambda)\,A^2(t)}\Bigg[a_1+a_3\,(a_1-2)\Big(a_3+(1-a_3)\,\mathrm{sech}^2[a_0\,x+c_6]\Big)\Bigg],
  \end{array}
\end{equation}

\begin{equation}\label{u55}
  \begin{array}{ll}
\beta^2(x,t)=\dfrac{1}{3\,(\lambda-1)\,A^2(t)}\Big[K_1+K_2\,A^{-2\,a_1}(t)+K_3\,\mathrm{sech}^2[a_0\,x+c_6]\Big],
  \end{array}
\end{equation}
where
\begin{equation}\label{u55-1}
\left\{
  \begin{array}{ll}
K_1=2\,a_5\,\big[a_3^2\,(a_1-2)^2+a_1\big]\big[\lambda(a_1+1)+\lambda-1\big],\\
\\
K_2=c_8\,(\lambda-1)\,\big[a_1\,(a_1+4)-a_3^2\,(a_1-2)^2\big],\\
\\
K_3=a_1\,a_0^2\,(a_3-1)(a_1-2)\big[\lambda(3a_3+1)+a_3-1\big].
  \end{array}
\right.
\end{equation}

\textbf{For the model (\ref{u314-2})}

\begin{equation}\label{u52-9}
  \begin{array}{ll}
\rho(x,t)=\dfrac{a_1\,a_5}{\chi\,(1-\lambda)\,A^2(t)}\Bigg[a_1+a_3\,(a_1-2)\Big(a_3+(1-a_3)\,\mathrm{sec}^2[a_0\,x+c_6]\Big)\Bigg],
  \end{array}
\end{equation}

\begin{equation}\label{u53-9}
  \begin{array}{ll}
p(x,t)=\dfrac{\lambda\,a_1\,a_5}{\chi\,(1-\lambda)\,A^2(t)}\Bigg[a_1+a_3\,(a_1-2)\Big(a_3+(1-a_3)\,\mathrm{sec}^2[a_0\,x+c_6]\Big)\Bigg],
  \end{array}
\end{equation}

\begin{equation}\label{u55-9}
  \begin{array}{ll}
\beta^2(x,t)=\dfrac{1}{3\,(\lambda-1)\,A^2(t)}\Big[K_1+K_2\,A^{-2\,a_1}(t)+K_3\,\mathrm{sec}^2[a_0\,x+c_6]\Big],
  \end{array}
\end{equation}
where
\begin{equation}\label{u55-1-9}
\left\{
  \begin{array}{ll}
K_1=2\,a_5\,\big[a_3^2\,(a_1-2)^2+a_1\big]\big[\lambda(a_1+1)+\lambda-1\big],\\
\\
K_2=c_8\,(\lambda-1)\,\big[a_1\,(a_1+4)-a_3^2\,(a_1-2)^2\big],\\
\\
K_3=a_1\,a_0^2\,(1-a_3)(a_1-2)\big[\lambda(3a_3+1)+a_3-1\big].
  \end{array}
\right.
\end{equation}

It is worth noting that the magnetic permeability is a variable quantity of $x$ and $t$. From equation (\ref{u213}), we can get the magnetic permeability for the models (\ref{u314-1}) and (\ref{u314-2}), respectively as the form:

\begin{equation}\label{u54}
  \begin{array}{ll}
\bar{\mu}(x,t)=\Big(\dfrac{\chi}{a_6^2\,a_0^2\,(a_3-1)}\Big)\,F^2_{12}(x)\,A^{-2\,a_2}(t)\,\mathrm{cosh}^{3-a_3}[a_0\,x+c_6],
  \end{array}
\end{equation}
and
\begin{equation}\label{u54-9}
  \begin{array}{ll}
\bar{\mu}(x,t)=\Big(\dfrac{\chi}{a_6^2\,a_0^2\,(1-a_3)}\Big)\,F^2_{12}(x)\,A^{-2\,a_2}(t)\,\mathrm{cos}^{3-a_3}[a_0\,x+c_6],
  \end{array}
\end{equation}
where the electro-magnetic field $F_{12}$ in these models is an arbitrary function of $x$ only.\\

For the line element (\ref{u314-1}) and (\ref{u314-2}), using equations (\ref{u22-1}), (\ref{u22-5}) and (\ref{u22-6}),
 we have the following physical properties: The volume element is
\begin{equation}  \label{u56}
V\,=\,
\left\{
  \begin{array}{ll}
\,a_6\,a_8\,A^{a_1+1}(t)\,\mathrm{cosh}^{a_3}[a_0\,x+c_6],\\
\\
\,a_6\,a_8\,A^{a_1+1}(t)\,\mathrm{cos}^{a_3}[a_0\,x+c_6].
\end{array}
\right.
\end{equation}

The expansion scalar, which determines the volume behavior of the
fluid, is given by:
\begin{equation}\label{u57}
  \begin{array}{ll}
\Theta=\dfrac{(1+a_1)\,\sqrt{a_5+c_8\,A^{-2\,a_1}}}{A(t)}.
  \end{array}
\end{equation}
The non-vanishing components of the shear tensor, $\sigma_i^j$, are:
\begin{equation}\label{u58}
  \begin{array}{ll}
\sigma_1^1=\dfrac{(2-a_1)\,\sqrt{a_5+c_8\,A^{-2\,a_1}}}{3\,A(t)},
  \end{array}
\end{equation}

\begin{equation}\label{u59}
  \begin{array}{ll}
\sigma_2^2=\dfrac{(a_1-2)\,(1+3a_3)\,\sqrt{a_5+c_8\,A^{-2\,a_1}}}{6\,A(t)},
  \end{array}
\end{equation}

\begin{equation}\label{u510}
  \begin{array}{ll}
\sigma_3^3=\dfrac{(a_1-2)\,(1-3\,a_3)\,\sqrt{a_5+c_8\,A^{-2\,a_1}}}{3\,A(t)},
  \end{array}
\end{equation}

\begin{equation}\label{u511}
  \begin{array}{ll}
\sigma_4^4=-\dfrac{2\,(1+a_1)\,\sqrt{a_5+c_8\,A^{-2\,a_1}}}{3\,A(t)}.
  \end{array}
\end{equation}
Hence the shear scalar $\sigma$, is given by:
\begin{equation}\label{u512}
  \begin{array}{ll}
\sigma^2=\Big(\dfrac{20+4\,a_1+11\,a_1^2+9\,a_3^2\,(2-a_1)^2}{36}\Big)\Big(\dfrac{a_5+c_8\,A^{-2\,a_1}}{A^2(t)}\Big).
  \end{array}
\end{equation}
The model does not admit acceleration and rotation, since $\dot{u}_{i}=0$ and $\omega_{ij}=0$. We can see that
\begin{equation}\label{u513}
\left\{
  \begin{array}{ll}
\dfrac{\sigma^1_1}{\Theta}=\dfrac{c_1}{3},\,\,\,\,\,\,\,\,\,\,\,\,\,\,\,\,\,\,\,\,\,\,\,\,\,\,\,\,\,\,\,
\,\,\,\,\,\dfrac{\sigma^2_2}{\Theta}=-\dfrac{c_1\,(1+3\,a_3)}{6},\\
\\
\dfrac{\sigma^3_3}{\Theta}=\dfrac{c_1\,(3\,a_3-1)}{6},\,\,\,\,\,\,\,\,\,\,\dfrac{\sigma^4_4}{\Theta}=-\dfrac{2}{3}.
  \end{array}
\right.
\end{equation}
where $a_1=\dfrac{2-c_1}{1+c_1}$. We found also, that
\begin{equation}\label{u514}
  \begin{array}{ll}
\dfrac{\sigma}{\Theta}=\dfrac{\sqrt{20+4\,a_1+11\,a_1^2+9\,a_3^2\,(2-a_1)^2}}{6\,(1+a_1)},
  \end{array}
\end{equation}
which means that the model does not approach isotropy for large limit $t$.

\begin{figure}[htbp]
    \centering
        \includegraphics[scale=.3]{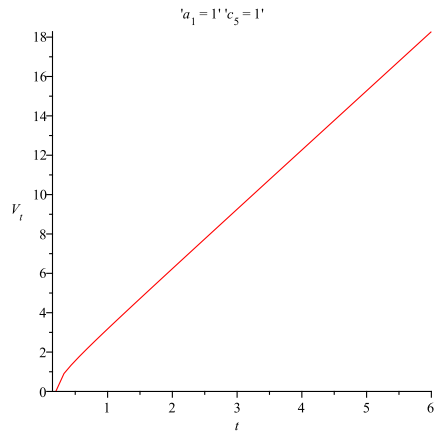}
        \caption{The temporal behaviour of three  space volume for $a_1=1$}
   \label{fig:fig2}
\end{figure}

The deceleration parameter is given by \cite{dec}
\begin{equation}\label{u514-1}
  \begin{array}{ll}
\mathbf{q}=-3\,\Theta^2\,\Big(\Theta_{;i}\,u^{i}+\dfrac{1}{3}\,\Theta^2\Big).
  \end{array}
\end{equation}
For the line element (\ref{u21}) and from (\ref{u22-5}), we have
\begin{equation}\label{u514-2}
  \begin{array}{ll}
\Theta_{;i}\,u^{i}=\dfrac{\partial\Theta}{\partial
t}=\dfrac{C_{tt}}{C}+\dfrac{B_{tt}}{B}-\dfrac{C_{t}^2}{C^2}-\dfrac{B_{t}^2}{B^2}
-\dfrac{\dot{A}^2}{A^2}+\dfrac{\ddot{A}}{A}.
  \end{array}
\end{equation}
For the line element (\ref{u314-1}) or (\ref{u314-2}) and from
(\ref{u57}), we have
\begin{equation}\label{u514-3}
  \begin{array}{ll}
\mathbf{q}=(a_1+1)^3\,A^{-4(a_1+1)}\,\big[c_8+a_5\,A^{2\,a_1}\big]\,\Big[2\,c_8\,(a_1+1)+a_5\,(a_1-2)\,A^{2\,a_1}\Big],
  \end{array}
\end{equation}
where $A$ is a function of $t$ satisfies the equation in (31).

\begin{figure*}[thbp]
\begin{tabular}{rl}
\includegraphics[width=6.5cm]{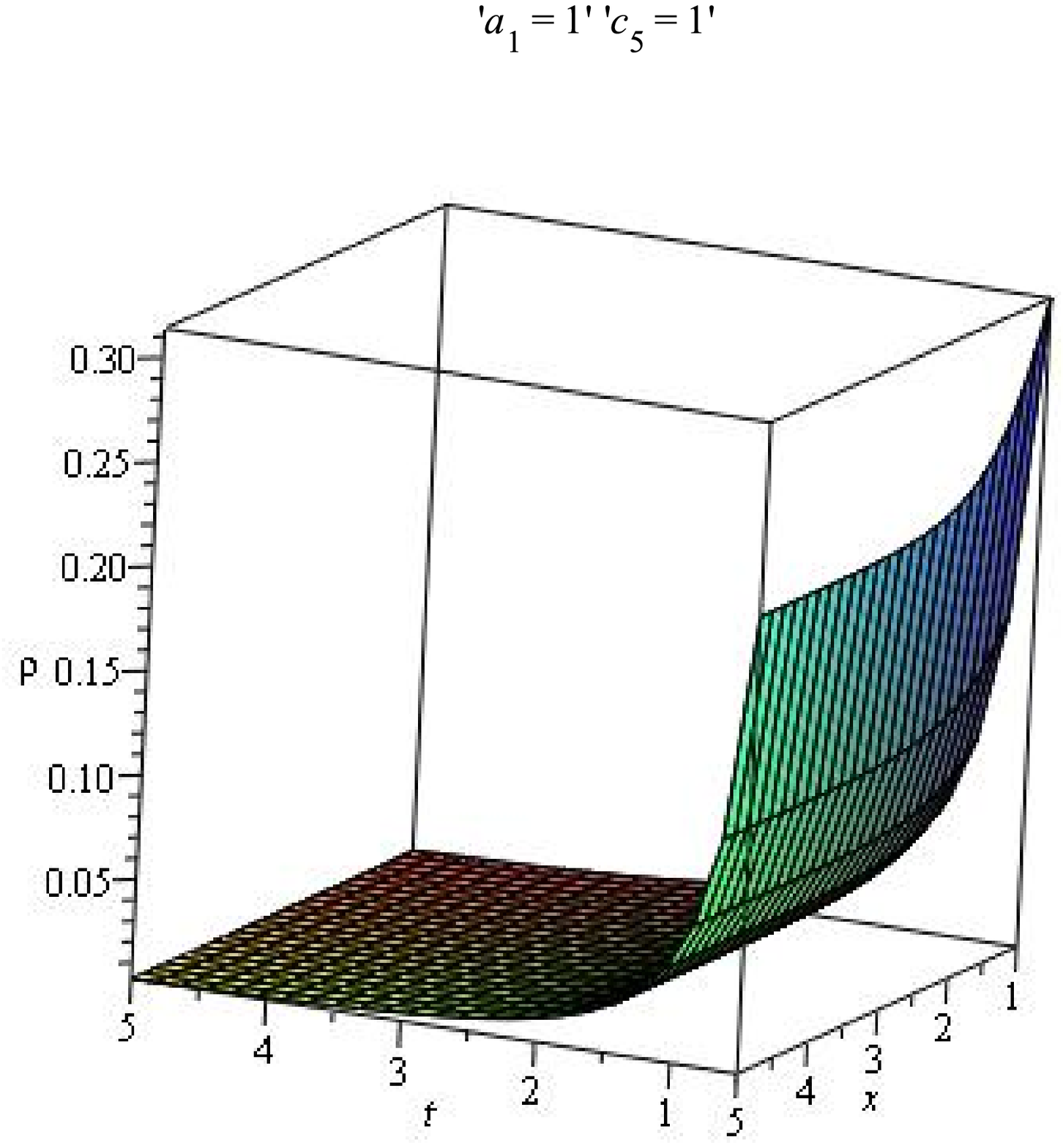}&
\includegraphics[width=6.5cm]{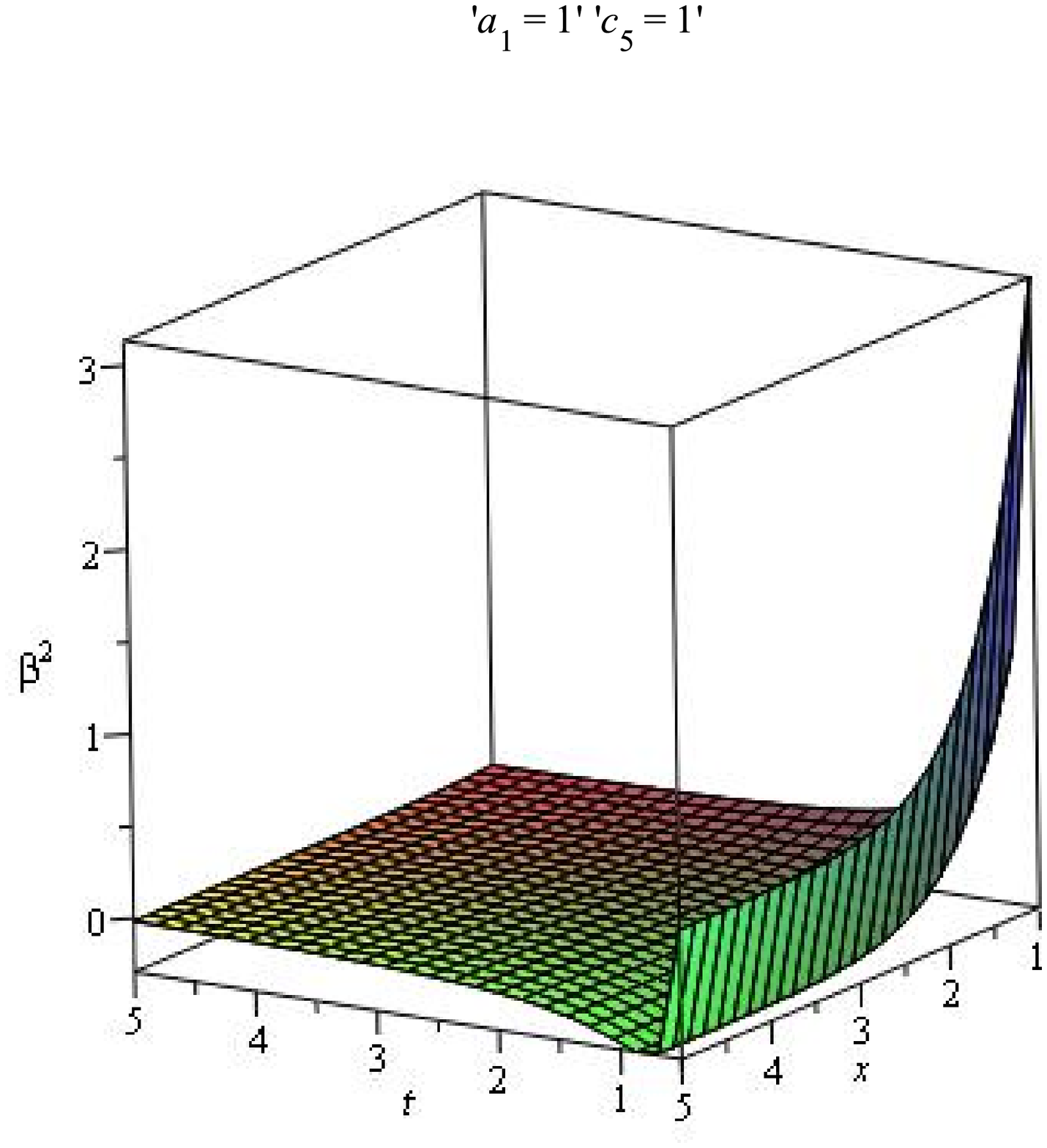}
\\
\end{tabular}
\caption{ (\textit{Left})  The variation of energy density with
respect to space and time  for $a_1=1$. (\textit{Right}) The
variation of energy displacement parameter  with respect to space
and time  for $a_1=1$.}
\end{figure*}


\begin{figure}[htbp]
    \centering
        \includegraphics[scale=.3]{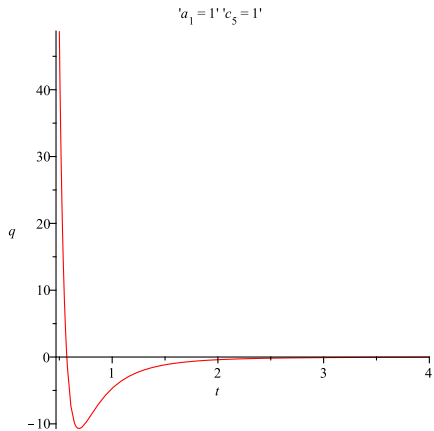}
        \caption{The deceleration parameter is shown with respect to time  for $a_1=1$}
   \label{fig:fig2}
\end{figure}

\pagebreak
\section{Concluding remarks}

We have investigated an inhomogeneous Bianchi type-I cosmological
model of the universe.    We have solved the modified Einstein
field equations within the framework of Lyra geometry.   In this
solution, we take magnetic field and perfect fluid together as the
source of gravitational field. We were able to investigate
simultaneously two types of spatial behavior of the space-time.
However, we have obtained five sets of solutions for the temporal
behavior of the space-time. One can notice that the solution set-1 gives the power law solutions of the metric potential.
Solution set-3 and set-4 give singularity-free solutions of
the universe. Solution set-3 indicates the proper volume
remain constant. Therefore, this case is not interesting. For
solution set-5, we could not get simple expression of coefficient metric $A$ in
terms of 't' rather a transcend form and consequently conclusion
can not be drawn easily. For the sake of brevity, we will discuss
various properties of the solution set two. Here, we observe that
the initial epoch will be $t = t_0 = \frac{\sqrt{c_8}}{a_0^2} -
b_2$. The model starts with an initial singularity with  $V
\rightarrow 0 $, while $\theta, \sigma^2$ diverge. In fact, it is
a point singularity as all the metric coefficients are zero at
this epoch. The temporal behavior of the proper volume is show in
figure 1. This indicates that after the initial singularity the universe  expands indefinitely.
 We have shown graphically the space and time variation of
 energy density and displacement parameter ( see figures 2 Left and Right), respectively. It is to be noted that displacement vector will not exist after infinite time.
We have calculated the  deceleration parameter as it  serves as an
indicator   whether the model accelerates. It is known that if  $q\,>\,0$
  the cosmological model decelerates whereas for $q\,<\,0$ the model
  accelerates. Recent observations   on supernova due to the
High-$z$ Supernova Search Team (HZT) and the Supernova Cosmology
Project (SCP) \cite{Riess1998, Perlmutter1998} confirm  that the
present expanding Universe is getting gradual acceleration.
Cosmologists argued  that  the expansion of the universe changed
from decelerating to accelerating. Figure 3 of our model confirms
this. Therefore, our model is very much realistic in the sense
that  at least theoretically it explains the recent experimental
findings through the Supernova Cosmology Project. One can assume
that displacement vector plays the role of additional energy
density, which causes the acceleration of the universe.



\begin{thebibliography}{99}

\bibitem{ali1} A. T. Ali, J. Comp. Appl. Math. \textbf{235} (2011) 4117.

\bibitem{bali11} R. Bali and N. K. Chandnani, J. Math. Phys. \textbf{49} (2008) 032502.

\bibitem{bali12} R. Bali and N. K. Chandnani, Int. J. Theor. Phys. \textbf{48} (2009) 1523.

\bibitem{bali13} R. Bali and U. K. Pareek, Astrophysics Space Sci. \textbf{312} (2007) 305.

\bibitem{bali14} R. Bali and R. Vadhwani, Int. J. Phys. Sci. \textbf{26(6)} (2011) 6172.

\bibitem{casama1} R. Casama, C. Melo and B. Pimentel, Astrophysics Space Sci. \textbf{305} (2006) 125.

\bibitem{book} S. S. De and F. Rahaman, Finsler geometry of hadrons and Lyra geometry: Cosmological aspects, Lambert Academic Publishing, Germany, 2012.

\bibitem{elsa1} M. F. El-Sabbagh and A. T. Ali, Int. J. Nonlinear Sci. Numer. Simulat. \textbf{6(2)} (2005) 151.

\bibitem{elsa2} M. F. El-Sabbagh and A. T. Ali, Commun. Nonlinear Sci. Numer. Simulat. \textbf{13} (2008) 1758.

\bibitem{dec}  A. Feinstein and J. lbanez, Class. Quantum Grav. \textbf{10} (1993) L227.

\bibitem{hal} W. D. Halford, Aust. J. Phys. \textbf{23} (1970) 863.

\bibitem{ha} E. R. Harrison, Phys. Rev. Lett. \textbf{30} (1973) 188.

\bibitem{katore1} S. D. Katore, R. S. Rane and K. S. Wankhade, Pramana J. Phys. \textbf{76(4)} (2011) 543.

\bibitem{AK} A. Krasinski, In homogeneous Cosmological Models, Cambridge University Press, Cambridge 1997.

\bibitem{kumar1} S. Kumar and C. P. Singh, Int. J. Mod. Phys. A \textbf{23} (2008) 813.

\bibitem{lich1} A. Lichnerowicz, Relativistic Hydrodynamics and Magneto-hydro-dynamics, W A Benjamin Inc. New-York, p.93, (1967).

\bibitem{ly} G. Lyra, Math. Z. \textbf{54} (1951) 52.

\bibitem{mel} M. A. Melvin, Ann. New York Acad. Sci. \textbf{262} (1975) 253.

\bibitem{Perlmutter1998} S. Perlmutter et al., Nature {\bf 391} (1998) 51.

\bibitem{prad1} A. Pradhan and P. Mathur, Fizika B. \textbf{18} (2009) 243.

\bibitem{prad21} A. Pradhan, I. Aotemshi and G. P. Singh, Astrophysics Space Sci. \textbf{288} (2003) 315.

\bibitem{prad22} A. Pradhan and A. K. Vishwakarma, J. Geom. Phys. \textbf{49} (2004) 332.

\bibitem{prad23} A. Pradhan and S. S. Kumhar, Astrophysics Space Sci. \textbf{321} (2009) 137.

\bibitem{prad23-2} A. Pradhan and P. Ram, Int. J. Theor. Phys. \textbf{48} (2009) 3188.

\bibitem{prad23-3} A. Pradhan, H. Amirhashchi and H. Zainuddin, Int. J. Theor. Phys. \textbf{50} (2011) 56.

\bibitem{prad23-4} A. Pradhan, A. Singh and R. S. Singh, Rom J. Phys. \textbf{56} (2011) 297.

\bibitem{prad24} A. Pradhan and A. K. Singh, Int. J. Theor. Phys. \textbf{50} (2011) 916.



\bibitem{rahaman1} F. Rahaman, B. Bhui and G. Bag, Astrophysics Space Sci. \textbf{295} (2005) 507.

\bibitem{rao1} V. U. M. Rao and T. Vinutha and M. V. Santhi, Astrophysics Space Sci. \textbf{314} (2008) 213.

\bibitem{rayc1} A. K. Raychaudhuri, Theoritical Cosmology, Oxford, p.80, (1979).

\bibitem{Riess1998} A. G. Riess et al, Astronomical J. {\bf 116} (1998) 1009.

\bibitem{saman1} G. C. Samanta and S. Debata, J. Mod. Phys. \textbf{3} (2012) 180.

\bibitem{sen1} D. K. Sen, Phys. Z. \textbf{149} (1957) 311.

\bibitem{SD} D. K. Senand K. A. Dunn, J. Math. Phys. \textbf{12} (1971) 578.

\bibitem{yadav1} A. K. Yadav, A. Pradhan and A. Singh, Rom. J. Phys. \textbf{56} (2011) 1019.

\bibitem{zel} Ya. B. Zeldovich, A. A. Ruzmainkin and D. D. Sokoloff, Magnetic field in Astrophysics, Gordon and Breach, New York (1993).

\bibitem{zia1} R. Zia and R. P. Singh, Rom. J. Phys. \textbf{57} (2012) 761.

\end{thebibliography}
\end{document}